\documentclass[11pt]{article}

\usepackage{authblk}
\usepackage{algorithm}
\usepackage{algorithmic}
\usepackage{graphicx}
\usepackage[margin=1in]{geometry}
\usepackage{color}
\usepackage{amsmath,amsfonts,amssymb}
\usepackage[title]{appendix}

\newcommand{\br}{\mathbf{r}}
\newcommand{\bk}{\mathbf{k}}

\begin{document}
	
	\title{Understanding the morphotropic phase boundary of\\ perovskite solid solutions as a frustrated state}
	
	\author[1]{Ying Shi Teh}
	\author[2]{Jiangyu Li}
	\author[1]{Kaushik Bhattacharya\footnote{Corresponding Author: bhatta@caltech.edu}}
	
	\affil[1]{California Institute of Technology, Pasadena CA 91125, USA}
	\affil[2]{Southern University of Science and Technology, Shenzhen, PRC}
	%\date{}
	
	\maketitle

{\bf	
%[250 words] 

Perovskite solid solutions that have a chemical composition A(C$_x$D$_{1-x})$O$_3$ with transition metals C and D substitutionally occupying the B site of a perovskite lattice are attractive in various applications for their dielectric, piezoelectric and other properties.  A remarkable feature of these solid solutions is the \emph{morphotropic phase boundary} (MPB), the composition across which the crystal symmetry changes.  Critically, it has long been observed that the dielectric and piezoelectric as well as the ability to pole a ceramic increases dramatically at the MPB.  While this has motivated much study of perovskite MPBs, a number of important questions about the role of disorder remain unanswered.    We address these questions using a new approach based on the random-field Ising model with long-range interactions that incorporates the basic elements of the physics at the meso-scale.  We show that the MPB emerges naturally in this approach as a frustrated state where stability is exchanged between two well-defined phases.  Specifically, long-range interactions suppress the disorder at compositions away from MPB but are unable to do so when there is an exchange of stability.   Further, the approach also predicts a number of experimentally observed features like the fragmented domain patterns and superior ability to pole at the MPB.  The insights from this model also suggest the possibility of entirely new materials with strong ferroelectric-ferromagnetic coupling using an MPB.
}

{\it \paragraph{Significance}
%[120 words]  
Perovskites are widely used in capacitor, ultrasonic, photonic, sensor and actuator applications for their dielectric, piezoelectric and optical properties.  Many of these properties are enhanced in perovskite solid solutions at compositions close to the morphotropic phase boundary (MPB).   This observation drives the search for new materials including lead-free piezoelectrics; yet the physics of MPB are incompletely understood.  We present a new approach which shows that the MPB arises as a frustrated state in a competition between local chemical disorder and long-range interactions.  It explains various experimental observations and the insights are useful in the search for new piezoelectrics.  Further, the model also suggests the possibility of entirely new phenomena by exploiting MPBs.
}
\vspace{\baselineskip}

\begin{figure}[t!]
	\begin{center}
		\includegraphics[width=0.9 \textwidth]{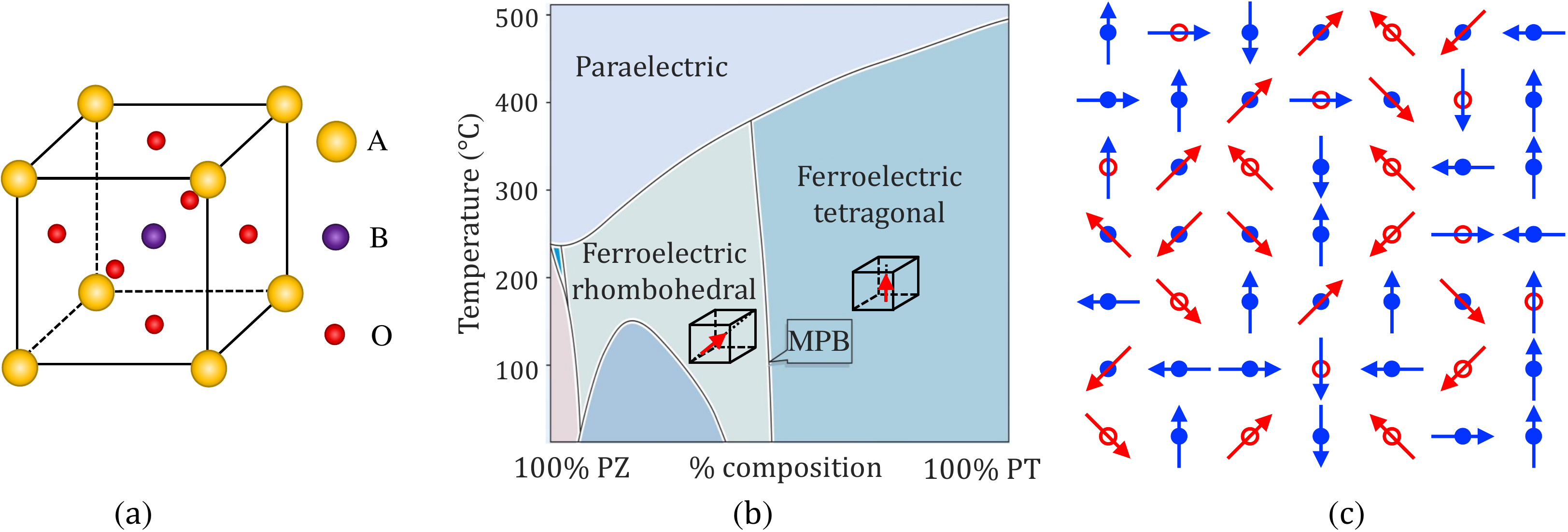}
	\end{center}
	\caption{(a) Perovskite structure.  (b) Phase diagram of PZT adapted from Cross \cite{c_nat_04}. (c) Schematic illustration of the lattice model proposed in this work.}
		\label{fig:back}
\end{figure}

\noindent  Perovskites are a class of materials with a chemical composition of ABO$_3$, where A and B are typically transition metals, and a crystal structure similar to that of the mineral perovskite CaTiO$_3$ shown in Figure \ref{fig:back}(a) (e.g. \cite{uchino_2009,tilley_2016}). It is common that these materials undergo a series of displacive phase transitions from a high temperature cubic ideal perovskite (Pm$\overline{3}$m) structure to distorted tetragonal (P4mm), rhombohedral (R3m), orthorhombic (Amm2) and other structures.  These lower symmetry structures may be non-centro-symmetric and therefore can become electrically polarized or magnetized.  Therefore these materials are widely used in capacitor, ultrasonic, optical, sensor and actuator applications for their dielectric and piezoelectric properties.

The basic structure is extremely stable and it is possible to have solid solutions A(C$_x$D$_{1-x})$O$_3$ where two metallic species C and D substitutionally occupy the B site of the lattice.  The low temperature structure in these compounds depends on composition.  A remarkable feature that is observed in a number of such solid solutions is the {\it morphotropic phase boundary} (MPB), the composition across which the symmetry changes, see Figure \ref{fig:back}(b).  This composition is largely independent of temperature (up until the temperature where it transforms to the cubic structure).  Critically, it has long been observed that the dielectric and piezoelectric as well as the ability to pole a ceramic increases dramatically at the MPB \cite{c_nat_04,Damjanovic2005,li_natmat_05}.   This has been central to the widespread use of lead zirconate titanate (PbZr$_x$Ti$_{1-x}$O$_3$  or PZT) that has a MPB at  $x= 0.52$ with a ferroelectric rhombohedral (R3m) structure in the Zr-rich compositions\footnote{PZT shows a second  rhombohedral (R3c) phase at low temperature at high Zr compositions, but we focus on the compositions near the MBP} and ferroelectric tetragonal (P4mm) structure in the Ti-rich phases.  The search for lead-free dielectric and piezoelectric materials has also focussed on solid solutions with MPBs (e.g. \cite{wu_2020} for a recent review).

Given the importance of MPBs, it has been and continues to be the subject of intense study.   Classically, it was believed that the tetragonal and rhombohedral phases coexist at the MPB.  This was challenged by the discovery of a low-symmetry monoclinic phase (Cm) at the MPB by Noheda {\it et al.} \cite{Noheda1999} using x-ray powder diffraction.  This was supported by first principles calculations that developed a composition-dependent hybrid pseudopotential \cite{Bellaiche2000}.  Importantly, it was recognized that the presence of a bridging phase enable a larger intrinsic piezoelectric effect at the MPB \cite{Fu2000,Damjanovic2005,Cao2004}.  Further, either the coexistence or the availability of a low symmetry bridging phase enabled a high extrinsic piezoelectric effect at the MPB \cite{li_natmat_05}.

Since then, there have been a number of studies of the crystal structure of MPB-PZT, and there are observations consistent with various structures.  Examples include the combination of two monoclinic phases (Cm and Ic, \cite{Cox2005} or Cm and Pm  \cite{Zhang2018}), combination of tetragonal (P4mm) and monoclinic (Cm) \cite{Ragini2002} and combination of rhombohedral (R3m) and monoclinic (Cm) \cite{Gorfman2011}.  This uncertainty has been attributed to the disorder in the composition resulting in a disorder in the structure, and the difficulty of resolving local structures \cite{noheda_rev}.   This role of disorder is also supported by first principles calculations \cite{Grinberg2002,Grinberg2004}.  This however raises the question as to why the disorder does not affect the structure away from the MPB.
Another interesting observation concerns the domain patterns.  Classical well-defined domain patterns are observed away from the MPB, but highly fragmented domain patterns are observed near the MPB \cite{Woodward2005}.  It has been argued that this fragmented domain pattern also contributes to the high piezoelectric response near the MPB \cite{Theissmann2007}.

In short, critical questions remains open.  Why is the effect of compositional disorder suppressed to form an unambiguous structure away from the MPB, but suddenly revealed at  the MPB?  Is there a definitive crystal structure at the MPB?  Why do domain patterns become fragmented near the MPB?  Does compositional disorder play a role in the ease of poling at the MPB?  Can the MPB be exploited to create new phenomena?  These questions are important because the MPB is the focus of the development of new materials.  However, they have proved to be challenging.  The disordered nature of the solid solution requires a large ensemble that takes it beyond the scope of direct first principles calculations without the introduction of an averaged pseudopotential.  On the other hand, phase-field Landau-Ginzburg methods can provide insight into domain patterns.  However, they are too coarse-grained to account for atomic-scale interactions and instead incorporate the MPB phenomenologically.

We address these questions using a new approach based on the random-field Ising model with long-range interactions that incorporates the basic elements of the physics at the meso-scale.  First, the B sites of a perovskite form a reference cubic lattice that is occupied randomly by atoms of either C or D species.  Second, the local quantum mechanical interactions create a propensity for the unit cell to break cubic symmetry depending on the species at the B site. Finally, there are long-range interactions due to ferroelectric, ferromagnetic and ferroelastic polarizations.  We create an effective Hamiltonian with these physics and study the ground states using the 
Markov chain Monte-Carlo (MCMC) method with cooling.  

In the first part of the paper, we show that this simple model provides new insights to the the questions concerning the MPB of ferroelectric solid solutions like PZT.  In particular, the long-range interactions which have an ordering feature overwhelm the local disorder in the C-rich and D-rich compositions with the exchange of stability taking place at a specific composition where the material is frustrated.  This frustration manifests itself as the MPB.  The frustrated state also enables easy poling as observed.  In the second part, we use the model to explore the possibility of obtaining materials with strong ferroelectric-ferromagnetic coupling using the insights obtained in the first part.  Such multi-ferroic coupling is limited in single materials \cite{hill}, and is thus realized using composite media.

%===================================================================================
\section{Ferroelectric solid solution}\label{section:model1}
%===================================================================================

\paragraph{Model}
Consider a $d$-dimensional periodic lattice ($d=2$ or $3$) with $N$ lattice points as shown in Figure \ref{fig:back}(c). Each lattice point $i$ is characterized by fixed (quenched) chemical composition ($c_i$) of either type $C$ ($c_i=0$ indicated by a red open circle in Figure \ref{fig:back}(c)) or type $B$ ($c_i=1$ indicated by a blue closed circle).  Each lattice point carries a dipole state ($\boldsymbol{p}_i$ indicated by the arrows in Figure \ref{fig:back}(c)) that can take one of a number of orientations determined by the Hamiltonian 
\begin{align}\label{eqn:total energy}
W_{tot}(\{\boldsymbol{p}_i\}; \{c_i\}) 
&= \sum_{i=1}^{N} h_{loc}(\boldsymbol{p}_i; c_i) -\frac{J_e}{2} \sum_{<i,j>} \boldsymbol{p}_i \cdot \boldsymbol{p}_j 
+ D_e W_{dip} (\{\boldsymbol{p}_i\}) 
- \mathbf{E}_{ext} \cdot \sum_{i}^{N} \boldsymbol{p}_i.
\end{align} 
The first term encodes the information that lattice site of type $C$ (respectively $D$) energetically prefers the set of dipole states $\mathcal{C}$ indicated by the red arrows (respectively $\mathcal{D}$ indicated by the blue arrows), though they can take states in $\mathcal{D}$  (respectively $\mathcal{C}$) with an energetic cost $h>0$:
\begin{equation}\label{eqn:local energy}
h_{loc}(\boldsymbol{p}_i;c_i) = 
\begin{cases}
	0 \quad \text{if $c_i=0$ and } \boldsymbol{p}_i \in \mathcal{C},  \text{ or, $c_i=1$ and } \boldsymbol{p}_i \in \mathcal{D}, \\
	h \quad \text{otherwise.}
\end{cases}
\end{equation}
The second term is the exchange energy with $J_e>0$ (and the sum is limited to nearest neighbors) that promotes like neighbors.  
The third term is the long-range electrostatic dipole-dipole interaction where
\begin{equation} \label{eq:dip}
W_{dip} (\{\boldsymbol{p}_i\}) = \frac{1}{(d-1) }\sum_{i,j=1}^{N} \sum_{\mathbf{R }}^{,}
\frac{1}{r_{ij}^d} \left[ \boldsymbol{p}_i \cdot \boldsymbol{p}_j - \frac{d (\boldsymbol{p}_i \cdot \br_{ij}) (\boldsymbol{p}_j \cdot \br_{ij})}{r_{ij}^2} \right] + \frac{2\pi}{d} \sum_{i=1}^{N} |\boldsymbol{p}_i|^2
\end{equation}
with strength $D_e$ which incorporates the dipole strength, lattice constant and electro-magnetic constants.  The final term is the influence of the applied external electric field $\mathbf{E}_{ext}$.

Given a lattice where the composition of each site is randomly assigned subject to a fixed average, we use a Markov chain Monte-Carlo (MCMC) method with cooling to obtain the equilibrium distribution at a given temperature.  The state is initialized by randomly assigning a polarization from $\mathcal{C}\cup\mathcal{D}$.  Adapting the Metropolis-Hastings algorithm to our multi-state setting, a site is chosen at random and its dipole state is updated to one of the $N_{states}$ states according to the transition probability
\begin{equation}\label{eqn:transition probability}
P_s = \frac{ \exp(-\beta W_{tot}^{(s)})}{\sum_{r=1}^{N_{states}}\exp (-\beta W_{tot}^{(r)})}, \qquad s = 1, 2, ... , N_{states}
\end{equation}
where $\beta$ is the inverse temperature and $N_{states}$ is the cardinality of $\mathcal{C}\cup\mathcal{D}$.  We avoid the system getting trapped in local minima at low temperatures by starting at a high temperature ($\beta=0$) and slowly cooling (increasing $\beta$) to the temperature of interest, while performing enough MCMC steps to reach equilibrium at each temperature. 
The details are provided in Methods.

\paragraph{Results}
We study an example motivated by PZT though the results are generic.  Here the C lattice points represent unit cells containing Zr atoms while the D lattice points represent unit cells containing Ti atoms.  Recall that the former prefers rhombohedral or $\langle 111 \rangle$ polarization states while the latter prefer tetragonal or $\langle 100 \rangle$ polarization states.  We begin in two dimensions $d=2$ so that the $\mathcal{C} = 1/\sqrt{2} \{  [1,1], [1,-1],[-1,1], [-1,-1]\}$  while $\mathcal{D} = \{ [1,0], [0,1], [-1,0], [0,-1]\}$.  We set $h=J_e=D_e=1$,  Ewald parameters $\sigma=0.157$ and $M_{cut}=16$, and conduct our simulations on a $256^2$ lattice.  In each simulation, we begin with an inverse temperature of $\beta=0$, and repeatedly increase its value with a small step size of $\Delta \beta = 0.05$ until we reach $\beta = 5$.   At each temperature value, at least 2000 Monte Carlo (MC) sweeps (each sweep consists of $N= 256^2$ steps) are performed with a total of $\approx 2\times10^5$ sweeps.

%%%%%%%%%%%%%%%%%%%%
\begin{figure}[t!]
	\centering
	\includegraphics[width=0.95 \textwidth]{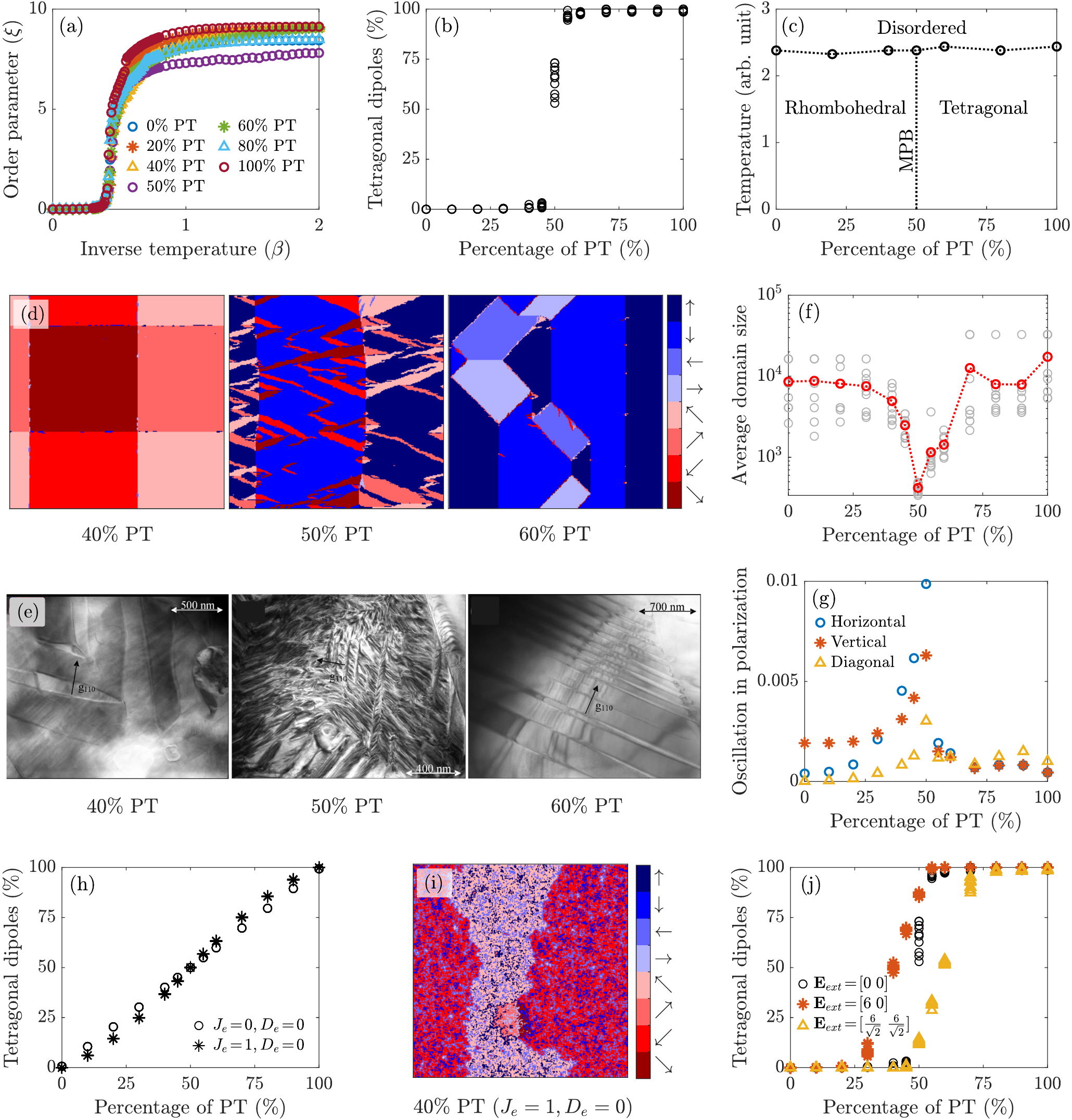}
	\caption{Emergence of a morphotropic phase transition (MPB) as a competition between short-range (compositional) disorder and long-range (exchange and electrostatic) interactions.  (a) Order parameter vs. inverse temperature for various compositions.  (b) Dipole orientation in the ordered phase at various compositions.  (c) Phase diagram showing the MPB.  (d) Domain patterns at various composition.  (e) Experimentally observed domain patterns at various compositions (reprinted with permission from Woodward, Knudsen, and Reaney \cite{Woodward2005}) (f) Average domain size vs. composition (average (red) and ten realizations (black)). (g) Small scale oscillations in terms of the finest Haar wavelet coefficient vs. composition.  (h) Domain pattern in the absence of long-range dipole-dipole interactions ($x=0.6$).  (i) Order parameter vs. inverse temperature in the absence of long-range dipole-dipole interactions. (j) The effect of external electric field and superior ability to pole at the MPB.  }
	\label{fig:mpb}
\end{figure}

Figure \ref{fig:mpb} shows the results of these simulations.  Figure \ref{fig:mpb}(a) shows the evolution of the order parameter ($\xi =  \frac{1}{\kappa_{max}} \sum_{\kappa=1}^{\kappa_{max}} \kappa C(\kappa)$ where $C(\kappa) = \langle \boldsymbol{p}_i \cdot \boldsymbol{p}_j \rangle $ is the correlation function over any two sites $i$ and $j$ that satisfy $\kappa-1<|\br_i-\br_j|\leq \kappa$) in a series of simulations with varying average composition.  The material is disordered at high temperature, but becomes ordered at low temperatures.  The phase transition is somewhat diffuse due to the disorder.   Figure \ref{fig:mpb}(b) shows the nature of the ordered phase.  Remarkably, we find that all dipoles are in the rhombohedral ($\mathcal C$) states till a critical composition of  about $50 \%$ beyond which all dipoles are in the tetragonal  ($\mathcal D$) states.  Indeed, at a composition of $33.3\%$, a third of the sites would prefer tetragonal dipoles.  However, the exchange and electrostatic interaction with the neighbors overwhelm this preference and instead force it into the rhombohedral state.  The opposite happens at a composition of $66.7\%$.  The exchange of stability between the rhombohedral and tetragonal states takes place at a well-defined critical composition.   This observation is extremely robust: Figure \ref{fig:mpb}(b) includes results from 10 realizations.  In short, we see the {\it emergence of the morphotropic phase boundary} (MPB).  The resulting phase diagram is shown in Figure \ref{fig:mpb}(c) (where the order-disorder transition temperature is taken to be the temperature corresponding to the maximum curvature of the $\xi-\beta$ curve).

The phase diagram is qualitatively consistent with experimental observations with the paraelectric phase at high temperature and different ordered phase at low temperatures depending on composition.  The paraelectric/ferroelectric transformation temperature is constant, and the composition of the MPB  is at $50\%$ since we take the energetic penalty to be equal for both $\mathcal{C}$ states on the $D$ site and $\mathcal{D}$ states on the $C$ site.  If we had taken them to be different, say $h_{CD}$ for the $\mathcal{C}$ states on the $D$ site and $h_{DC}$ for the $\mathcal{D}$ states on the $C$ site, then a simple argument shows that the MPB would occur at a $D$ composition of $h_{DC}/(h_{CD} + h_{DC}$).  This difference would also make the paraelectric/ferroelectric transformation temperature composition dependent.

The resulting domain patterns are also interesting.  Figure \ref{fig:mpb}(d) shows the typical domain patterns at three different compositions (Animations of the simulation are provided in Supplementary Information (SI)).  We see the 2D analogs of $71^\circ$ or $107^\circ$ domain walls with $(10)$ normals at composition of $40\%$, and we see the 2D analog of $90^\circ$ domain walls with $(11)$ normals at composition of $60\%$.  However, at the MPB (composition $50 \%$), we see a highly fragmented and frustrated state which is a mixture of rhombohedral ($\mathcal C$) and tetragonal ($\mathcal D$) states with no clear domain pattern.  The domain walls no longer follow the typical low-order crystallographic directions.  Figure \ref{fig:mpb}(f) shows that the average domain size falls precipitously at the MPB compared to that at all other compositions.   Figure \ref{fig:mpb}(g) shows that the oscillations in polarization in the horizontal, vertical and diagonal directions are also magnified at the MPB.  Specifically, we take the Haar wavelet transform of the domain pattern and Figure \ref{fig:mpb}(g) shows the normalized sum of squares of the horizontal, vertical and diagonal detail coefficients obtained from level one (finest level) Haar wavelet decomposition averaged over 10 realizations. These observations are consistent with experimental observations.   Figure \ref{fig:mpb}(e) reproduces the experimental observations of Woodward, Knudsen, and Reaney \cite{Woodward2005}: classical well-defined domain patterns are observed away from the MPB, but highly fragmented domain patterns are observed near the MPB as in our simulations (Figure \ref{fig:mpb}(d)).

We comment that both the long-range interaction and the disorder in composition are necessary for this behavior.  In the absence of the long-range dipole-dipole interaction (i.e., when $D_e=0$), the average number of dipoles track the composition as shown in Figure \ref{fig:mpb}(h).  Further, while there is phase segregation when $J_e\ne0$, the domains are not structured as shown in Figure \ref{fig:mpb}(i).  Similarly, 
these complex domain patterns do not appear when the composition is not random (SI).

The situation is largely similar with some difference in detail in three dimensions (see SI). We observe that the MPB emerges with local and exchange energies  even in the absence of long-range dipole-dipole interactions (e.g. $h=J_e=1, D_e=0$).   This is consistent with the fact that local critical dimension of a random-field Ising model is 2.  In an Ising model with two states where $h$ is random and $J>0$, the lattice is always disordered in two dimensions  (i.e., does not undergo the order-disorder transition) while the lattice can be ordered in three dimension under random field $h$ of moderate strength at small enough temperatures \cite{aw_prl_89,n_rev_97}.   However, the domain patterns look more like those shown in Figure \ref{fig:mpb} (i) than actual ferroelectric domains in the absence of long-range interactions.   Our results are also consistent with the observation that dipolar interactions in an Ising model with two states leads to stripe domains in two dimensions \cite{sam_prb_96}.

Finally, we consider the effect of an applied external electric field in Figure \ref{fig:mpb}(j).   Under $[1 \ 0]$ field that makes the tetragonal ($\mathcal{B}$) dipoles energetically more favorable, we still see that the tetragonal ($\mathcal{B}$) dipoles continue to be suppressed for small Ti ($B$) concentrations.  However, tetragonal ($\mathcal{B}$) dipoles emerge gradually at a composition smaller than the MPB, increasing with composition till they become all tetragonal at a composition slightly larger than the MPB.  So, there is a transition with composition, but transition is gradual and not sharp as in the case of zero electric field.  There is a corresponding behavior when an electric field is applied in the $[1 \ 1]$ direction.  This shows that materials close to the MPB can undergo an electric-field-imposed rhombohedral-to-tetragonal transition close to the MPB but not at other compositions as experimentally observed \cite{Kovacova2014}.  The corresponding microstructures are shown in SI.

%===================================================================================
\section{A multiferroic material}\label{section:model2}
%===================================================================================

A number of perovskites are known to be ferromagnetic \cite{kundu}.  Therefore, we explore the possibility of creating a multiferroic material (one that shows a strong coupling between electric and magnetic polarization) by exploiting the insight that the competition between short-range disorder and long-range interaction can lead to unique ordering behavior.  Consider a solid solutions of two materials, $C$ which prefers the $\mathcal{C}$ states that are ferroelectrically polarized in one crystallographic direction but with no ferromagnetism, and a material $D$ that prefers the $\mathcal{D}$ states that are ferromagnetically polarized in a different crystallographic direction but with no ferroelectricity.  The states are also mechanically distorted with the spontaneous strain corresponding to their ferroelectric/ferrrogmagnetic directions.   Since perovskites readily form solid solutions, and since perovskites can be both ferroelectric and ferromagnetic, it is natural to look for such systems in this class of materials.

\paragraph{Model}
Consider a $d$-dimensional periodic lattice ($d=2$ or $3$) with $N$ lattice points. Each lattice point $i$ is characterized by fixed (quenched) chemical composition ($c_i$) of either type $C$ ($c_i=0$) or type $D$ ($c_i=1$).  The state $s_i$ at the $i^{th}$ lattice point is characterized by an electrical dipole $\boldsymbol{p}_i$, magnetic dipole $\boldsymbol{m}_i $ and an elastic strain $\boldsymbol{e}_i $, and governed by the Hamiltonian
\begin{align} \label{eq:fefm}
W_{tot}(\{s_i\};\{c_i\}) &= W_{loc}(s_i;c_i ) + W_{exc}(\{s_i\})  \\
& + D_eW_{dip}(\{\boldsymbol{p}_i\}) + D_m W_{dip}(\{\boldsymbol{m}_i\}) + D_s W_{str}(\{\boldsymbol{e}_i\}) -  \mathbf{E}_{ext} \cdot \sum_{i}^{N} \boldsymbol{p}_i -  \mathbf{H}_{ext} \cdot \sum_{i}^{N} \boldsymbol{m}_i, \nonumber
\end{align}
where 
\begin{equation}
h_{loc}(s_i;c_i) = 
\begin{cases}
	0 \quad \text{if $c_i=0$ and } s_i \in \mathcal{C}  \text{ or $c_i=1$ and } s_i \in \mathcal{D} \\
	h \quad \text{otherwise}
\end{cases}
\end{equation}
describes the local preference,
\begin{equation}
 W_{exc}(\{s_i\}) = -\frac{J_p}{2} \sum_{<i,j>} \mathbf{p}_i \cdot \mathbf{p}_j  -\frac{J_m}{2} \sum_{<i,j>} \mathbf{m}_i \cdot \mathbf{m}_j,
\end{equation}
is the exchange, $W_{dip}$ is given in (\ref{eq:dip}), $W_{str}$ is the strain energy (Methods) and the final two terms describe the role of the external electric and magnetic fields.
We find the equilibrium states at low temperatures using an MCMC method with cooling as before.

\begin{figure}[t!]
	\begin{center}
		\includegraphics[width=0.7 \textwidth]{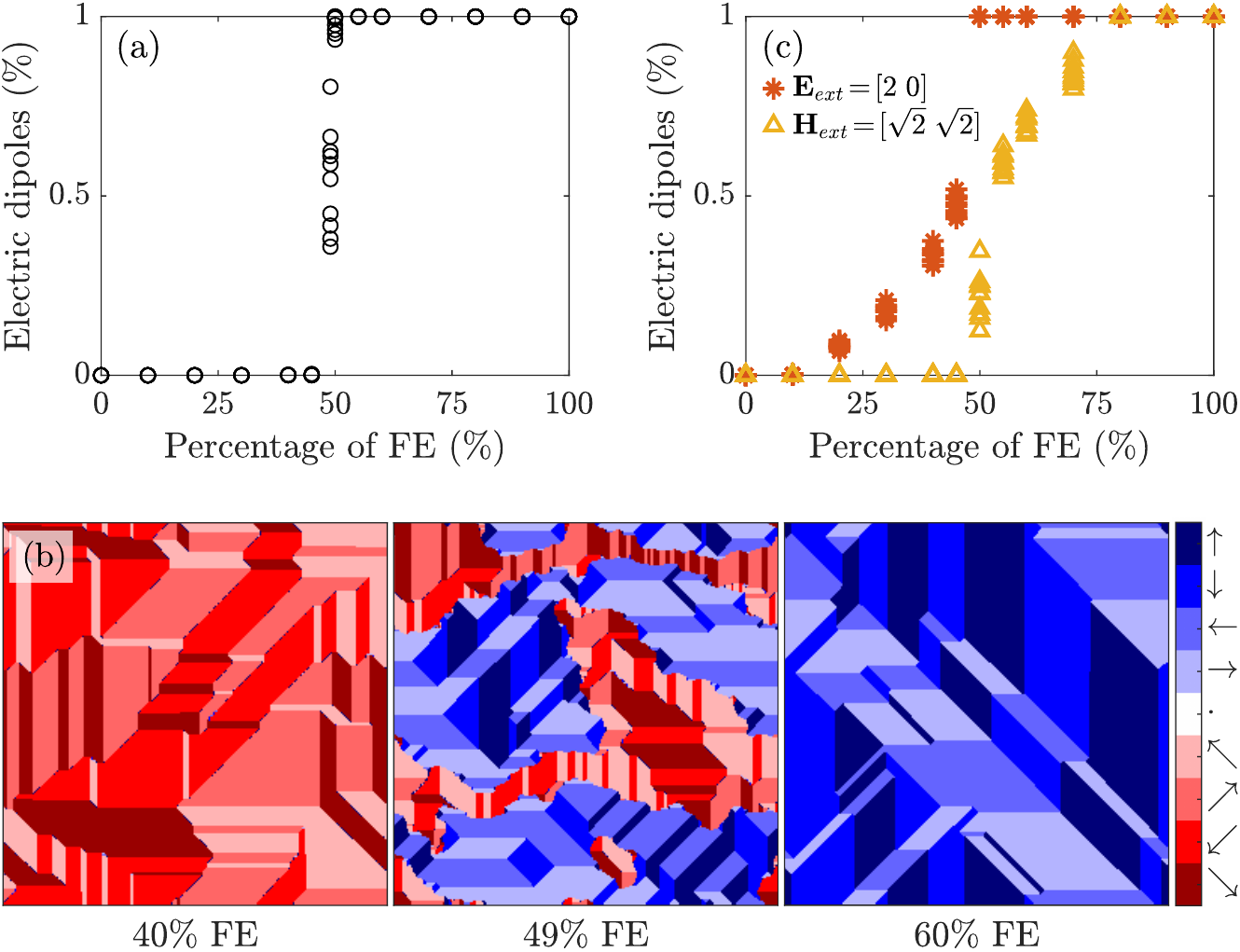}
	\end{center}
	\caption{Morphotropic phase boundary as means of creating a multiferroic material. (a) The ordered phase in a random lattice of a ferroelectric and ferromagnetic material shows a MPB between a ferroelectric phase and a ferromagnetic phase.  (b) Domain patterns at various compositions.  (c) The effect of applied field and a ferromagnetic/ferroelectric transformation.}
		\label{fig:FEFM}
\end{figure}

\paragraph{Results}
We consider two dimensions and assume that each lattice point can take one of nine states: four ferroelectric $\mathcal{C}$ states with $s_1 = \left\{ \mathbf{p}_1 = [1,0], \mathbf{m}_1 = 0, \mathbf{e}_1 = [[1,0],[0,-1]] \right\}$ and states $s_2$ though $s_4$ related to $s_1$ by symmetry; one zero state $s_5 = \left\{ \mathbf{p}_5 = 0, \mathbf{m}_5 = 0, \mathbf{e}_5 = 0 \right\}$; and four ferromagnetic $\mathcal{D}$ states with $s_6 = \left\{ \mathbf{p}_6 = 0, \mathbf{m}_6 = 1/\sqrt{2} [1,-1], \mathbf{e}_6 = [[0,1],[1,0]] \right\}$ and states $s_7$ though $s_9$ related by symmetry.  

Figure \ref{fig:FEFM} shows the results when $h = 2$, $J_e = J_m = D_e = D_m = D_s = 1$.  Once again, we have an order-disorder phase transition, and we observe the emergence of a MPB in the form of a sharp transition  at $49\%$ from a ferromagnetic phase at low $C$ compositions to a ferroelectric phase at high $C$ compositions in the absence of any external fields.  In particular, all ferroelectric $\mathcal{C}$ states are suppressed at low $C$ compositions, and all ferromagnetic $\mathcal{D}$ are suppressed at high $C$ compositions as shown in Figure \ref{fig:FEFM}(a).  Further, the zero state is always suppressed. Thus, there are no multiferroic states except at the MPB.  Furthermore, as shown in  Figure \ref{fig:FEFM}(b), we have classical domain walls at low and high compositions but fragmented non-classical domain walls at the MPB.

Figure \ref{fig:FEFM}(c)  shows the effect of external electric and magnetic fields.  The application of external magnetic field leads to a ferroelectric-to-ferromagnetic phase transition while the application of an external electric field leads to a ferromagnetic to ferroelectric phase transition at compositions close the MPB.  In other words, we have a strongly coupled multiferroic material close to the MPB.

\newpage
%===================================================================================
\section*{Methods} \label{app:dipole-dipole energy}
%===================================================================================

\paragraph{Dipole-dipole energy}
There are two important steps in the accurate and efficient computation of the dipole-dipole energy.  First, the expression (\ref{eq:dip}), in particular the first term, is conditionally convergence and we adopt  Ewald summation \cite{Ewald1921,Cerda2008,Wang2019} to separate it into a short-range contribution that is calculated in real space ($W_{dip}^r$), a long-range contribution that is calculated in Fourier space ($W_{dip}^k$), a self energy ($W_{dip}^{k\_self}$) and a surface term ($W_{dip}^{surf}$) that depends on the boundary condition.  Specifically, we rewrite the first term of (\ref{eq:dip}) as
\begin{equation}
W_{dip}^0 = W_{dip}^{r} + W_{dip}^{k} - \frac{1}{d (2\pi)^{d/2-1} \sigma^d } \sum_{i=1}^{N} |\boldsymbol{p}_i|^2 + \frac{2^{d-1}\pi}{(d-1)dV} \left| \sum_{i=1}^{N} \boldsymbol{p}_i \right|^2,
\end{equation}
where
\begin{eqnarray*}
W_{dip}^r &=& -\frac{1}{2} \sum_{i,j=1}^{N} \sum_{\mathbf{R}}^{,}  (\nabla_{\br_i} \cdot \boldsymbol{p}_i )(\nabla_{\br_j} \cdot \boldsymbol{p}_j) G_r(|\br_i-\br_j-\mathbf{R}|), \\
W_{dip}^{k} &=&
-\frac{1}{2V}
\sum_{\substack{\bk \neq \mathbf{0} \\ \bk \in \mathbb{K}^d}}
|\widetilde{\boldsymbol{p}}(\bk) \cdot i\bk|^2 \widetilde{G}_{\sigma}(\bk),
\end{eqnarray*}
with
$$
G_r(r) = \begin{cases}
\text{Ei} \left( -\frac{r^2}{2\sigma^2} \right), & d=2 \\
-\frac{1}{r} \text{erfc} \left( \frac{r}{\sqrt{2} \sigma} \right), & d=3
\end{cases}, \quad
\widetilde{G}_\sigma(\bk) = -\frac{4\pi}{k^2} \exp (-k^2 \sigma^2/2), 
$$
$\widetilde{\boldsymbol{p}}(\bk)$ the discrete Fourier transform of $\boldsymbol{p}$, $\mathbb{K}^d$ the Brillouin zone, and $\sigma$  a parameter chosen to be sufficiently small such that the $W_{dip}^{r}$ term is negligible but large enough to keep the calculation of $W_{dip}^{k}$ kept tractable.

Second, notice that for any flip, $\Delta W_{dip} \approx -\mathbf{E}_i \Delta \boldsymbol{p}_i$ where $\mathbf{E}_i = -\nabla_{\boldsymbol{p}_i} W_{dip}$ is the electric field.   While the conditional convergence means that $\mathbf{E}_i $ has to be recomputed after each flip, the error is small for individual flips.  Therefore we update $\mathbf{E}_i$ only every $\sqrt{N}$ flips where $N$ is the size of the lattice.  We can then perform $\sqrt{N}$ flips independently and in parallel thereby enabling acceleration on a graphical processing unit (GPU).  Further, $\mathbf{E}_i$ is readily computed using fast Fourier transforms which can also be implemented on GPUs.  

Specifically, the electric field at lattice point $\br_i$ due to other dipoles is
\begin{equation}
\mathbf{E}_i = -\nabla_{\boldsymbol{p}_i} W_{dip}^0 = \mathbf{E}_i^k - \mathbf{E}_i^{k\_self} + \mathbf{E}_i^{surf}.
\end{equation}
The difficult term is $\mathbf{E}_i^k$.   Due to the exponential decay in $\tilde{G}_{\sigma}(\bk)$, we may limit the summation over all $\mathbf{m} \in \mathbf{Z}^d$ to $\{ \mathbf{m} \in \mathbb{Z}^d, -M_{cut}/2\leq m_i<M_{cut}/2 \}$ and 
\begin{equation}
\mathbf{E}_i^k = \frac{4\pi}{V} \sum_{\bk \in \widetilde{\mathbb{M}}^d} \widetilde{\boldsymbol{p}}(\bk) \cdot
\underbrace{
	\left(
	\sum_{\mathbf{m} \in \mathbb{Z}^d} \mathbf{g}(\bk + 2\pi\mathbf{m})
	\right)}_{\mathbf{A}(\bk)}
\exp(i\bk \cdot \br_i),
\end{equation}
$$
\mathbf{g}(\bk_\mathbf{m}) = 
\begin{cases}
\mathbf{0} \quad &\text{if } \bk_\mathbf{m} = \mathbf{0}, \\
\bk_m \otimes \bk_m \widetilde{G}_{\sigma}(\bk_m) \quad &\text{otherwise}.
\end{cases}
$$
Note that we can precompute $\mathbf{A}(\bk)$ and exploit fast Fourier transform (FFT) to obtain an efficient algorithm:
(i) FFT to compute $\widetilde{\boldsymbol{p}}(\bk) $, (ii) multiplication with pre-computed $\mathbf{A}$ and (iii) inverse FFT to obtain $\mathbf{E}_i^k $ in real space.   See Algorithm \ref{alg:SAMC-GPU}.

\begin{algorithm}[t]
	\caption{GPU-accelerated computational method}\label{alg:SAMC-GPU}
	\begin{algorithmic}[1]
		\STATE Initialize dipole states $\{\boldsymbol{p}_i\}_{i=1}^{N}$
		\STATE Initialize $\beta = 0$
		\WHILE {$\beta < \beta_{max}$}
		\FOR {iteration $=1,\dotsc,\theta_{max}$}
		\STATE Construct $\{\boldsymbol{p}_i\}_{i=1}^{N}$
		\STATE Perform FFT of $\{\boldsymbol{p}_i\}_{i=1}^{N}$ to obtain $\widetilde{\boldsymbol{p}}(\bk)$
		\STATE Perform pointwise multiplication with $\mathbf{A}$ tensor
		\STATE Perform inverse FFT to obtain $\{ \mathbf{E}_i^k \}_{i=1}^{N}$
		\STATE Determine $\{ \mathbf{E}_i^{surf} \}_{i=1}^{N}$ using parallel sum reduction, and compute the net electric field \\ $\mathbf{E}_i + \mathbf{E}_{ext}$
		\STATE Generate $\sqrt{N}$ number of lattice points at random and perform MC updates on these points in parallel \\
		with $\Delta W_{tot} = -(\mathbf{E}_i + \mathbf{E}_{ext}) \Delta \boldsymbol{p}_i + \Delta W_{dip}^{self} + \Delta W_{local} + \Delta W_{exchange}$
		\ENDFOR
		\STATE $\beta \leftarrow \beta + \Delta \beta$
		\ENDWHILE 
	\end{algorithmic}
\end{algorithm}

%===================================================================================
\paragraph{Strain energy} \label{app:strain energy}
%===================================================================================

The strain energy $W_{str}$ in (\ref{eq:fefm}) is taken to be the strain energy of a continuum region with transformation strain $\boldsymbol{e}^*(\br)$ and uniform isotropic elastic modulus characterized by Lam\'e constants $\lambda$ and $\mu$.  We set $\boldsymbol{e}^*(\br) =  \begin{bmatrix} e_1^*(\br) & e_2^*(\br) \\ e_2^*(\br) & -e_1^*(\br) \end{bmatrix}$ and assume that $\boldsymbol{e}^*(\br) $ is pixelated with $\boldsymbol{e}^* = \boldsymbol{e}_i$ in the pixel containing the i$^{th}$ lattice point.  We can show (SI) 
\begin{equation}
W_{str} = W_{str}^{k} + W_{str}^{c\_self},
\end{equation}
where
\begin{eqnarray*}
W_{str}^{k} &=& \frac{\mu}{V(1-\nu)} \sum_{\mathbf{k} \neq \mathbf{0}} \left(
B_{11}(\mathbf{k}) |\widetilde{e}_1^*(\mathbf{k})|^2 +
B_{22}(\mathbf{k}) |\widetilde{e}_2^*(\mathbf{k})|^2 + 
2 B_{12}(\mathbf{k}) \operatorname{Re} \left( \overline{\widetilde{e}_1^*(\mathbf{k})} \widetilde{e}_2^*(\mathbf{k}) \right) \right), \\
W_{str}^{c\_self} &=& \frac{\mu}{V(1-\nu)} \left( D_{11} \sum_{\alpha=1}^{N} |{e_1^*}^{\alpha}|^2 + D_{22} \sum_{\alpha=1}^{N} |{e_2^*}^{\alpha}|^2 + 2D_{12} \sum_{\alpha=1}^{N} {e_1^*}^{\alpha} {e_2^*}^{\alpha} \right),\\
B_{11}(\mathbf{k}) &=& \frac{(k_1^2 - k_2^2)^2}{k^4} \exp(-k^2 \sigma^2),\\
B_{12}(\mathbf{k}) &=& \frac{2k_1 k_2 (k_1^2 - k_2^2)}{k^4} \exp(-k^2 \sigma^2), \\
B_{22}(\mathbf{k}) &=& \frac{4k_1^2 k_2^2}{k^4} \exp(-k^2 \sigma^2)
\end{eqnarray*}
and $\widetilde{\boldsymbol{e}}^*(\mathbf{k})$ is the discrete Fourier transform of $\boldsymbol{e}^*$.  The constants $D_{11}$, $D_{22}$ and $D_{12}$ in the term are determined such that the total strain energy is zero for the homogeneous case.  %In the simulations, the constant factor $2\mu/(1-\nu)$ is disregarded since the material parameters can be varied through $D_s$ in (\ref{eq:fefm}).

\newpage
\paragraph{Author contributions}
YST performed all the simulations presented in this work.  All authors participated in the formulation of the model and analysis of the results.  YST and KB took the lead in drafting the manuscript which was finalized by all authors.

\paragraph{Acknowledgment}
YST and KB gratefully acknowledge useful discussions with Roubing Bai, as well as the support of the De Logi Foundation through a gift to the California Institute of Technology.

%===================================================================================
%\section{References}
%===================================================================================
\bibliographystyle{unsrt}
\bibliography{References}

\end{document}

\title{Supplementary Information for \\
		Understanding morphotropic phase boundary of \\ perovskite solid solutions as a frustrated state}
	
	\author[1]{Ying Shi Teh}
	\author[2]{Jiangyu Li}
	\author[1]{Kaushik Bhattacharya\footnote{Corresponding Author: bhatta@caltech.edu}}
	
	\affil[1]{California Institute of Technology, Pasadena CA 91125, USA}
	\affil[2]{Southern University of Science and Technology, Shenzhen, PRC}
	\date{}
	
	\maketitle

\newpage

\noindent
Movie M1: Evolution of domain patterns during the simulation process of Markov chain Monte Carlo with cooling.

\begin{figure}[H]
	\begin{center}
		\includegraphics[width=0.5 \textwidth]{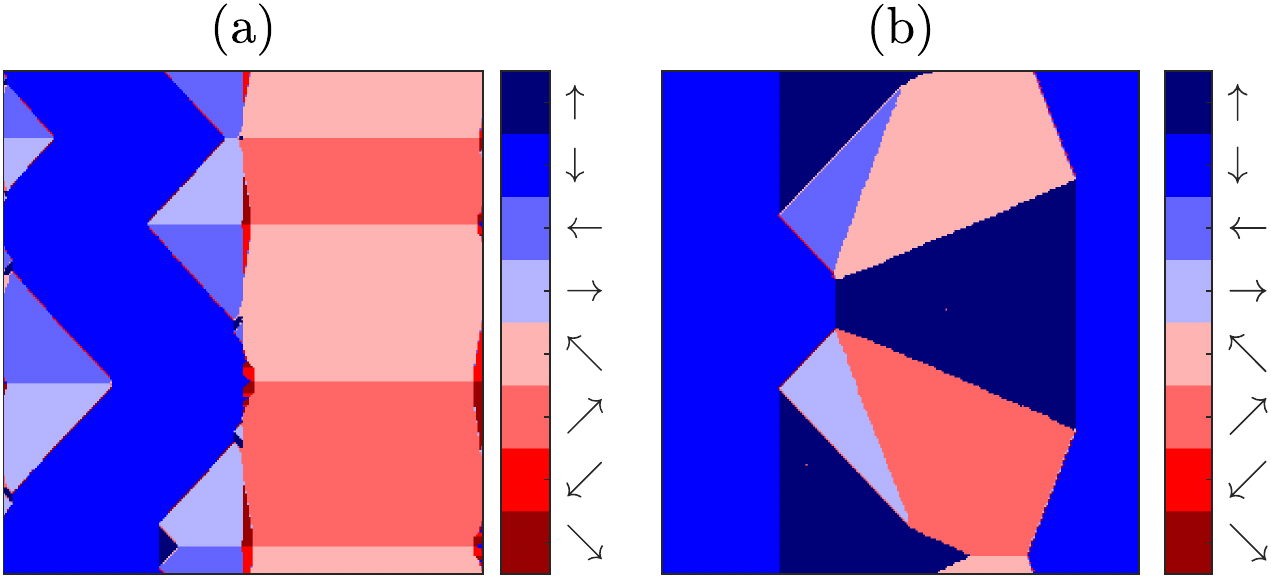}
	\end{center}
	\caption{Domain patterns obtained when (a) the composition is separated with Zr ($A$) sites on the left side and Ti (B) sites on the right side, and (b) the composition is regularly alternating (that is it has a checkerboard pattern). Note that complex domain patterns do not appear for these two cases when the composition is not random but has 50\% PT.}
		\label{fig:S1}
\end{figure}

\begin{figure}[H]
	\begin{center}
		\includegraphics[width=0.8 \textwidth]{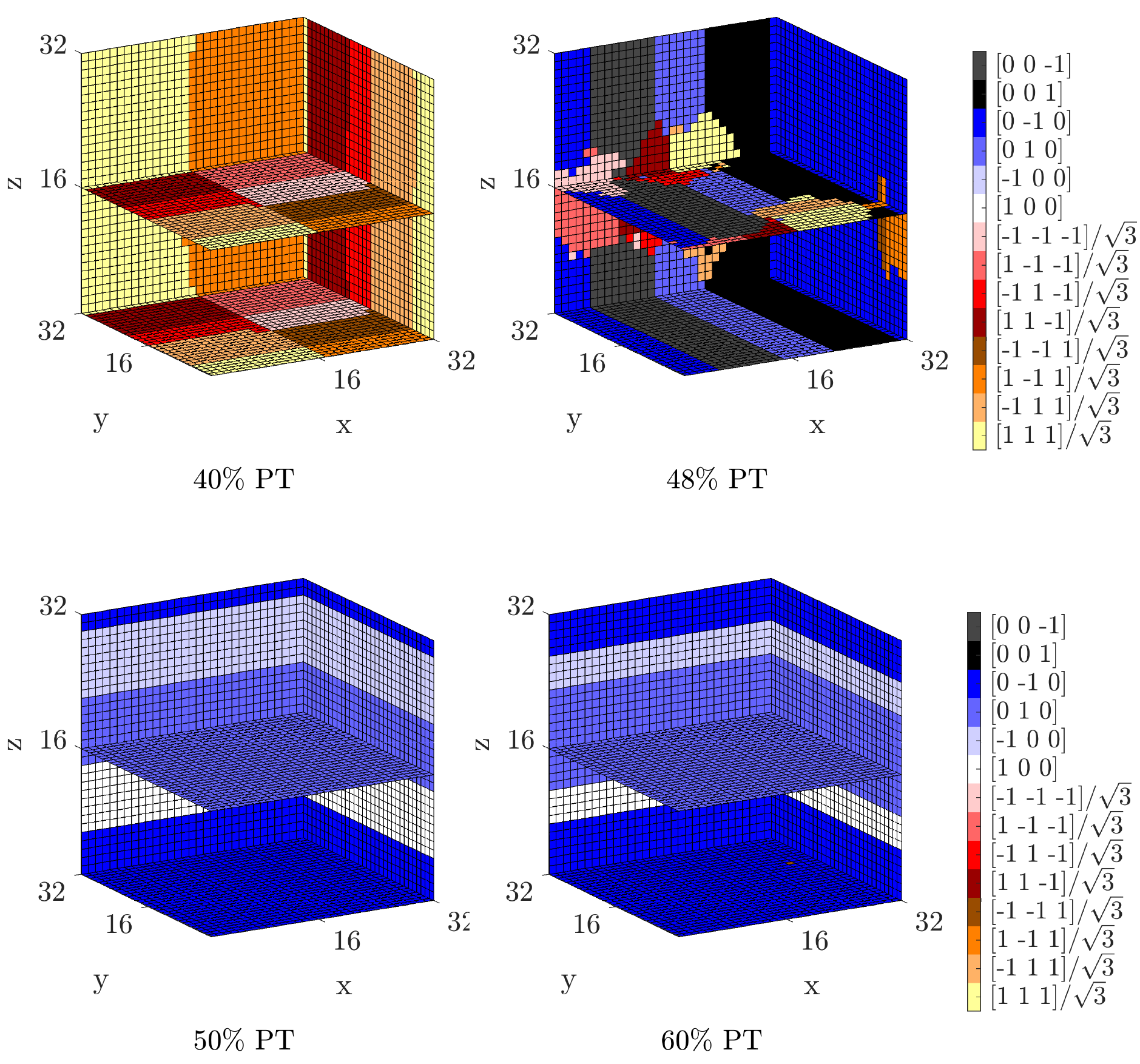}
	\end{center}
	\caption{Domain patterns at various compositions for 3D simulations ($h=1,J=0.5,D_e=1$). Similar to the 2D simulations, rhombohedral dipole states dominate at 40\% PT composition while tetragonal dipole states dominate at 60\% PT composition. At a composition of 48\%, we see the emergence of morphotropic phase boundary (MPB). The smaller-scale domain patterns at the MPB are less obvious in this case and this may be attributed to the size limit imposed by the significantly higher computational costs of the 3D simulations. }
	\label{fig:S2}
\end{figure}

\begin{figure}[H]
	\begin{center}
		\includegraphics[width=0.8 \textwidth]{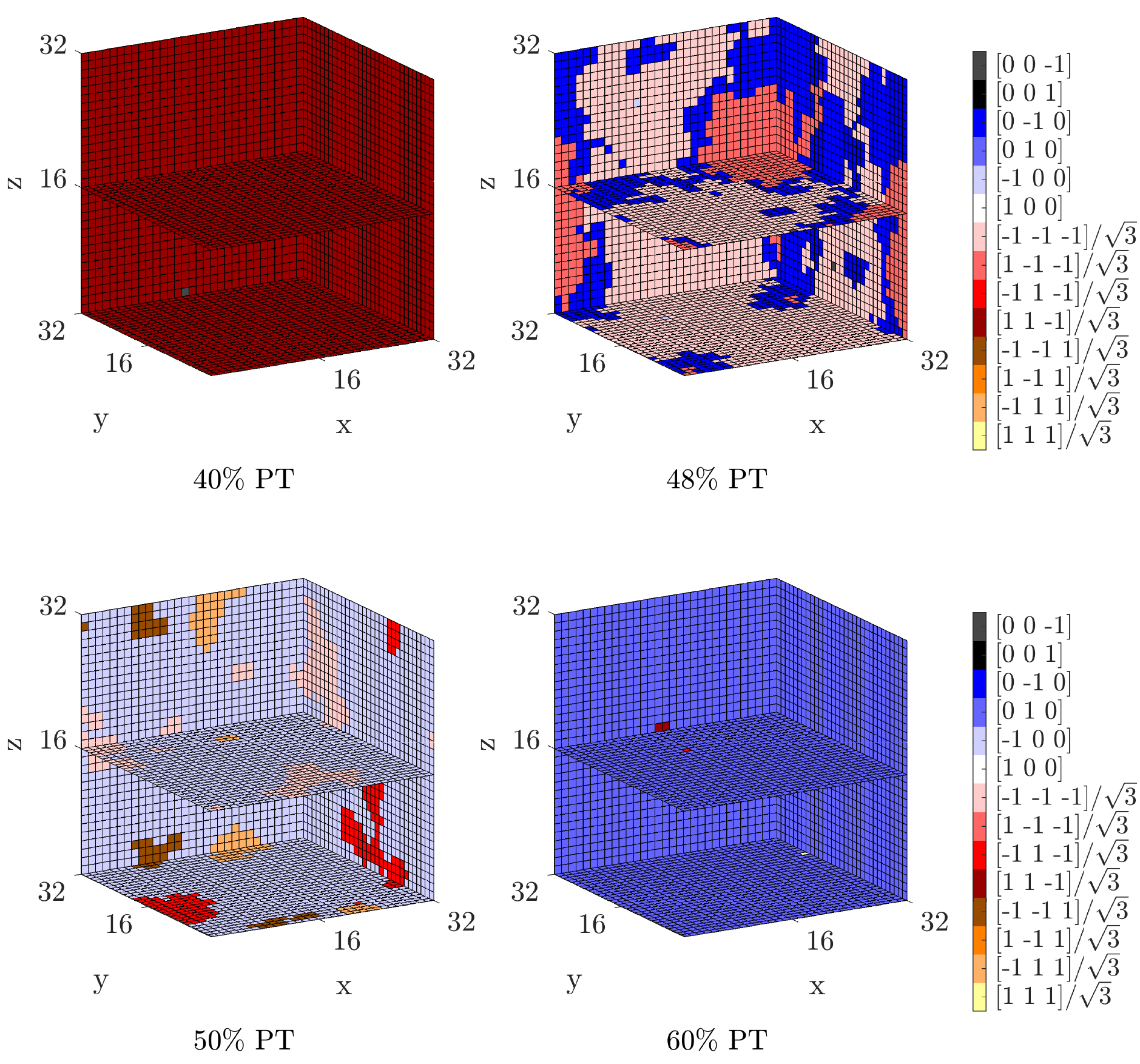}
	\end{center}
	\caption{Domain patterns at various compositions for 3D simulations in the absence of dipole-dipole interactions ($h=1,J=1,D_e=0$).}
	\label{fig:S2_extra}
\end{figure}

\begin{figure}[H]
	\begin{center}
		\includegraphics[width=0.7 \textwidth]{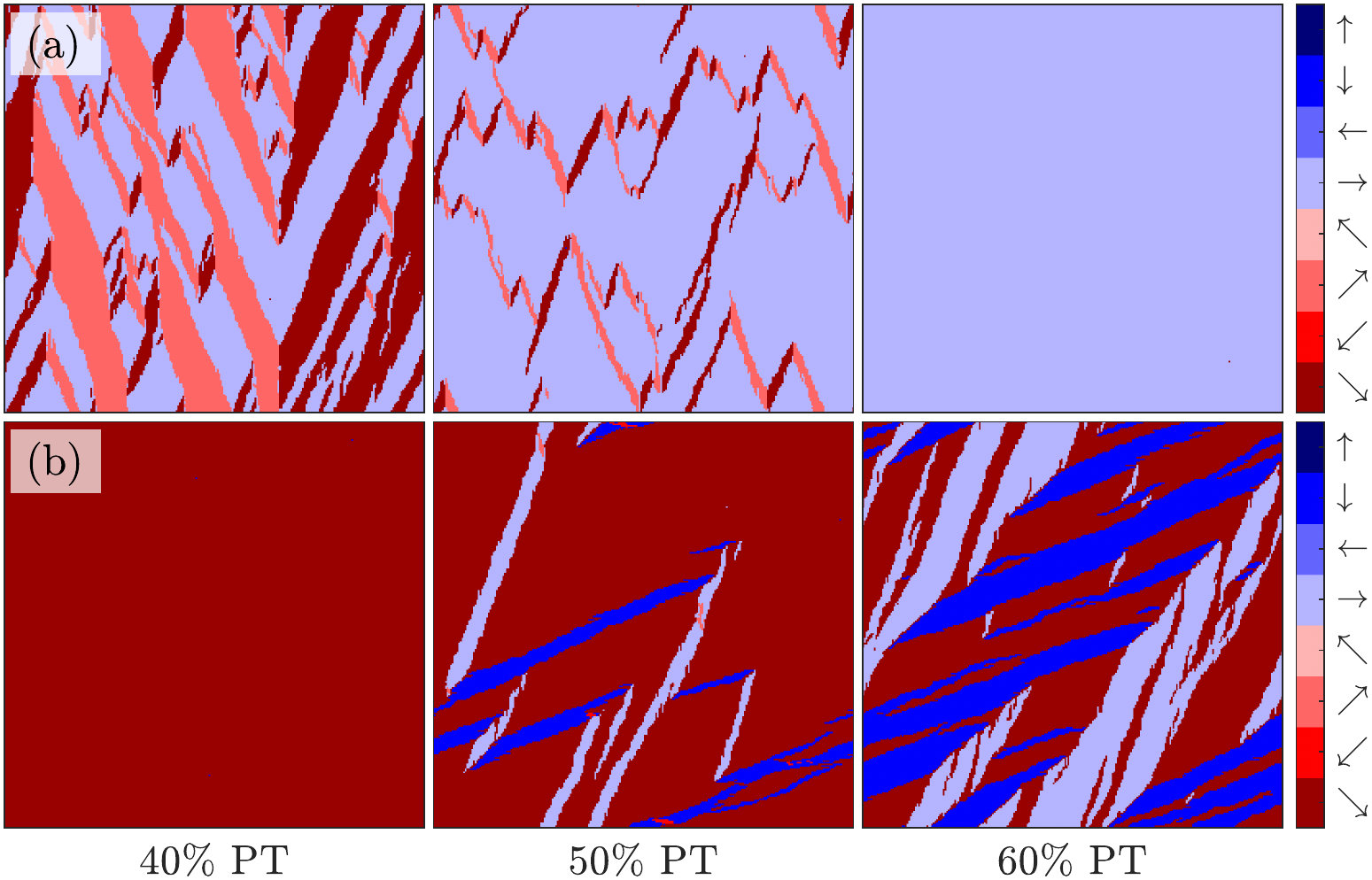}
	\end{center}
	\caption{Domain patterns at various composition under applied electric fields (a) $\mathbf{E}_{ext} = [6 \; 0]$ and (b) $\mathbf{E}_{ext} = [\frac{6}{\sqrt{2}} \; \frac{6}{\sqrt{2}}]$.}
	\label{fig:S3}
\end{figure}

\newpage
%===================================================================================
\section*{Strain energy expression}
%===================================================================================
Consider an infinite elastic body with a periodic volume $V$ that undergoes a field of transformation strain $e_{ij}^{*}(\mathbf{x})$. The transformation strain field is also periodic with $V$, i.e. $e_{ij}^{*}(\mathbf{x}) = e_{ij}^{*}(\mathbf{x}+\mathbf{R})$ for any translation vector $\mathbf{R}$. Using Einstein notation, the elastic strain energy stored in the periodic volume $V$ is given by

\begin{equation}\label{eqn:str1}
W_{str} = \frac{1}{2} \int_V C_{ijkl} \left( e_{ij}(\mathbf{x}) - e_{ij}^{*}(\mathbf{x}) \right) \left( e_{kl}(\mathbf{x}) - e_{kl}^{*} (\mathbf{x}) \right) d\mathbf{x} .
\end{equation}

Assume 2D plane strain, stress-free boundary condition in the infinity and homogeneous isotropic linear elastic material, i.e. $C_{ijkl} = \lambda \delta_{ij} \delta_{kl} + \mu (\delta_{ik} \delta_{jl} + \delta_{il} \delta_{jk})$, where $\lambda$ and $\mu$ are the Lam\'e constants. By solving the equilibrium equation $C_{ijkl} \frac{\partial}{\partial x_j}(e_{kl}(\mathbf{x}) - e^*_{kl}(\mathbf{x})) = 0$ in Fourier space and converting equation (\ref{eqn:str1}) into similar form using Parseval's theorem, we conclude 
\begin{equation}
W_{str} = W_1 + W_2 + W_3,
\end{equation}
\begin{equation}
W_1 = -\frac{1}{2 \mu V} \sum_{\substack{\mathbf{k} \neq \mathbf{0} \\ \bk \in \mathbb{K}^2}} \overline{\widetilde{M}_{ij} (\mathbf{k})} \widetilde{M}_{kl} (\mathbf{k}) \left[ \frac{k_j k_l}{k^2} \delta_{ik} - \frac{1}{2(1-\nu)} \frac{k_i k_j k_k k_l}{k^4} \right],
\end{equation}
\begin{equation}
W_2 = \frac{1}{2} \int_{V} C_{ijkl} e_{ij}^{*}(\mathbf{x}) e_{kl}^{*}(\mathbf{x}) d\mathbf{x} ,
\end{equation}
\begin{equation}
W_3 = -\frac{V}{2} C_{ijkl} \langle e_{ij}^{*} (\mathbf{x}) \rangle \langle e_{kl}^{*}(\mathbf{x}) \rangle,
\end{equation}
where $M_{ij}(\mathbf{x}) = C_{ijkl} e_{kl}^{*}(\mathbf{x})$ and $\widetilde{M}$ refers to the Fourier transform of $M$.

For our problem, it suffices to write $e^*(\mathbf{x})$ as
\begin{equation}
\boldsymbol{e}^{*}(\mathbf{x}) = 
\left[ {\begin{array}{cc}
	e_1^*(\mathbf{x}) & e_2^*(\mathbf{x}) \\
	e_2^*(\mathbf{x}) & -e_1^*(\mathbf{x}) \\
	\end{array} } \right].
\end{equation}
Then,
\begin{equation}
\widetilde{\mathbf{M}}^{*}(\mathbf{x}) = 2\mu 
\left[ {\begin{array}{cc}
	\widetilde{e}_1^*(\mathbf{k}) & \widetilde{e}_2^*(\mathbf{k}) \\
	\widetilde{e}_2^*(\mathbf{k}) & -\widetilde{e}_1^*(\mathbf{k}) \\
	\end{array} } \right]
\end{equation}
and
\begin{multline}
W_1 = -\frac{2\mu}{V} \sum_{\substack{\mathbf{k} \neq \mathbf{0} \\ \bk \in \mathbb{K}^2}} \left[ |\widetilde{e}_1^*(\mathbf{k})|^2 + |\widetilde{e}_2^*(\mathbf{k})|^2 \right] \\
+ \frac{\mu}{V(1-\nu)} \sum_{\substack{\mathbf{k} \neq \mathbf{0} \\ \bk \in \mathbb{K}^2}} \left[
\frac{(k_1^2 - k_2^2)^2}{k^4} |\widetilde{e}_1^*(\mathbf{k})|^2 +  \frac{4 k_1^2 k_2^2}{k^4} |\widetilde{e}_2^*(\mathbf{k})|^2 + \frac{4 k_1 k_2 (k_1^2 - k_2^2)}{k^4} \operatorname{Re} \left( \overline{\widetilde{e}_1^*(\mathbf{k})} \widetilde{e}_2^*(\mathbf{k}) \right) \right].
\end{multline}

The first term above cancels out with $W_2 + W_3$ since 
\begin{align*}
W_2 + W_3  &= 2\mu V \left[ \left\langle (e_1^*)^2 \right\rangle + \left\langle (e_2^*)^2 \right\rangle \right] - 2\mu V \left[ \langle e_1^* \rangle^2 + \langle e_2^* \rangle^2 \right] \\
&= \frac{2\mu}{V}  \sum_{\mathbf{k}} \left[ |\widetilde{e_1^*}|^2 + |\widetilde{e_2^*}|^2 \right] - 2\mu V \left[ \langle e_1^* \rangle + \langle e_2^* \rangle \right] \\
&=  \frac{2\mu}{V}  \sum_{\mathbf{k} \neq \mathbf{0}} \left[ |\widetilde{e_1^*}|^2 + |\widetilde{e_2^*}|^2 \right].
\end{align*}

Consider $N$ point inclusions, the field of transformation strain in the periodic volume $V$ is
\begin{equation}
e^*(\mathbf{x}) = \sum_{\alpha=1}^{N} {e^*}^{\alpha} \delta(|\mathbf{x} - \mathbf{x}^{\alpha}|).
\end{equation}
However, the Dirac delta function above introduces an issue when computing the self-energy. It also makes the sum in the Fourier space only conditionally convergent. Instead, we replace the delta function by a Gaussian function $g_{\sigma}(r)$. We then add a self-energy correction term $W_{str}^{c\_self}$, similar to the case of electrostatic energy or magnetostatic energy. The constants $D_{11}$ and $D_{22}$ in the term are determined such that the total strain energy is zero for the homogeneous case, that is when all the elastic dipoles are equal (${e^*}^{1} = {e^*}^{2} = \cdots = {e^*}^{N}$).

Finally, we have
\begin{equation}
W_{str} = W_{str}^{k} + W_{str}^{c\_self},
\end{equation}
where
$$
W_{str}^{k} = \frac{\mu}{V(1-\nu)} \sum_{\mathbf{k} \neq \mathbf{0}} \left[
B_{11}(\mathbf{k}) |\widetilde{e}_1^*(\mathbf{k})|^2 +
B_{22}(\mathbf{k}) |\widetilde{e}_2^*(\mathbf{k})|^2 + 
2 B_{12}(\mathbf{k}) \operatorname{Re} \left( \overline{\widetilde{e}_1^*(\mathbf{k})} \widetilde{e}_2^*(\mathbf{k}) \right) \right],
$$
$$
W_{str}^{c\_self} = \frac{\mu}{V(1-\nu)} \left[ D_{11} \sum_{\alpha=1}^{N} |{e_1^*}^{\alpha}|^2 + D_{22} \sum_{\alpha=1}^{N} |{e_2^*}^{\alpha}|^2 + 2D_{12} \sum_{\alpha=1}^{N} {e_1^*}^{\alpha} {e_2^*}^{\alpha} \right],
$$
$$
\widetilde{e}^*(\mathbf{k}) = \sum_{\alpha=1}^{N} {e^*}^{\alpha} \exp(-i\mathbf{k} \cdot \mathbf{x}^{\alpha}),
$$
$$
B_{11}(\mathbf{k}) = \frac{(k_1^2 - k_2^2)^2}{k^4} \exp(-k^2 \sigma^2),
$$
$$
B_{12}(\mathbf{k}) = \frac{2k_1 k_2 (k_1^2 - k_2^2)}{k^4} \exp(-k^2 \sigma^2),
$$
$$
B_{22}(\mathbf{k}) = \frac{4k_1^2 k_2^2}{k^4} \exp(-k^2 \sigma^2).
$$

Note that $W_{str}^{k}$ also has a self-energy component given by
$$
W_{str}^{k\_self} = \frac{\mu}{V(1-\nu)} \left[ C_{11} \sum_{\alpha=1}^{N} |{e_1^*}^{\alpha}|^2 + C_{11} \sum_{\alpha=1}^{N} |{e_2^*}^{\alpha}|^2 + 2 C_{12} \sum_{\alpha=1}^{N} {e_1^*}^{\alpha} {e_2^*}^{\alpha} \right]
$$
$$
C_{11} = \sum_{\mathbf{k} \neq 0} B_{11}(\mathbf{k})
$$
$$
C_{12} = \sum_{\mathbf{k} \neq 0} B_{12}(\mathbf{k})
$$
$$
C_{22} = \sum_{\mathbf{k} \neq 0} B_{22}(\mathbf{k})
$$
For easy computation of the change in energy due to the change in strain states, we introduce the following generalized stress values.
\begin{align*}
\sigma_1^{\alpha} &= \frac{\partial (W_{str}^{k} - W_{str}^{k\_self})}{\partial {e_1^*}^{\alpha}} \\
&= \frac{2\mu}{V(1-\nu)} \sum_{\mathbf{k} \neq \mathbf{0}} 
\operatorname{Re} \left\{ \left[
B_{11}(\mathbf{k}) \widetilde{e}_1^*(\mathbf{k}) + B_{12}(\mathbf{k}) \widetilde{e}_2^*(\mathbf{k})
\right] \exp(i\mathbf{k} \cdot \mathbf{x}^{\alpha}) \right\}
- \frac{2\mu}{V(1-\nu)} (C_{11} {e_1^*}^{\alpha} + C_{12} {e_2^*}^{\alpha}),
\end{align*}
\begin{align*}
\sigma_2^{\alpha} &= \frac{\partial (W_{str}^{k} - W_{str}^{k\_self})}{\partial {e_2^*}^{\alpha}} \\
&= \frac{2\mu}{V(1-\nu)} \sum_{\mathbf{k} \neq \mathbf{0}} 
\operatorname{Re} \left\{ \left[
B_{12}(\mathbf{k}) \widetilde{e}_1^*(\mathbf{k}) + B_{22}(\mathbf{k}) \widetilde{e}_2^*(\mathbf{k})
\right] \exp(i\mathbf{k} \cdot \mathbf{x}^{\alpha}) \right\}
- \frac{2\mu}{V(1-\nu)} (C_{12} {e_1^*}^{\alpha} + C_{22} {e_2^*}^{\alpha}).
\end{align*}
The change in strain energy can then be approximated as 
\begin{equation}
\Delta W_{str} \approx \sum_{\alpha=1}^{N} (\sigma_1^{\alpha} \Delta {e_1^*}^{\alpha} + \sigma_2^{\alpha} \Delta {e_2^*}^{\alpha}) + \Delta W_{str}^{k\_self} + \Delta W_{str}^{c\_self}.
\end{equation}
%In the main text, $\boldsymbol{e}^*$ and lattice vectors are normalized while $2\mu/(1-\nu)$ is replaced by a dimensionless constant $S$.